\documentclass[a4paper,10pt]{article}
\usepackage{graphicx,xcolor,geometry,url,multicol}
\usepackage[super,numbers,sort&compress]{natbib}
\setcitestyle{open={},close={},citesep={,}}
\usepackage{amssymb,amsmath}
\usepackage[T1]{fontenc}
\usepackage[font=small,labelfont=bf]{caption}

\usepackage{floatrow}

\usepackage{titlesec}
\titlespacing*\section{0pt}{12pt plus 4pt minus 2pt}{2pt plus 2pt minus 2pt}
\titlespacing*\subsection{0pt}{12pt plus 4pt minus 2pt}{2pt plus 2pt minus 2pt}
\titlespacing*\subsubsection{0pt}{4pt plus 2pt minus 2pt}{2pt plus 2pt minus 2pt}

\usepackage[bookmarks=true,breaklinks=true,hypertexnames=false,bookmarksnumbered=true]{hyperref}
 \hypersetup{
pdfauthor = { Adam R.~H.~Stevens },
pdftitle = { Aus Astro CO2 Paper },
pdfsubject = { Astronomy },
pdfkeywords = { -- },
pdfcreator = {LaTeX with hyperref package},
pdfproducer = {pdflatex},
colorlinks,
    linkcolor={red!50!black},
    citecolor={red!50!black},
    urlcolor={blue!50!black}}

\makeatletter
\newcommand*{\myfnsymbolsingle}[1]{%
  \ensuremath{%
    \ifcase#1% 0
    \or % 1
      *%   
    \or % 2
      \dagger
    \or % 3  
      \ddagger
    \or % 4   
      \mathsection
    \or % 5
      \mathparagraph
    \or \wedge \or \# \or \forall \or \odot \or \oplus \or \$ \or \pounds
    \else % >= 6
      \@ctrerr  
    \fi
  }%   
}   
\makeatother

\usepackage{alphalph}
\newalphalph{\myfnsymbolmult}[mult]{\myfnsymbolsingle}{}

\begin{document}

\begin{center}
{\LARGE \bf The imperative to reduce carbon emissions in astronomy}\\[0.4cm]
\today\\[0.1cm]
Accepted for publication in Nature Astronomy \\[0.4cm]
\end{center}
{\Large Adam R.~H.~Stevens$^{1,2\dagger}$, Sabine Bellstedt$^1$, Pascal J.~Elahi$^{1,2}$ and Michael T.~Murphy$^3$}\\[0.2cm]
{\small $^1$International Centre for Radio Astronomy Research, The University of Western Australia, Crawley, WA 6009, Australia\\
$^2$Australian Research Council Centre of Excellence for All Sky Astrophysics in 3 Dimensions (ASTRO 3D)\\
$^3$Centre for Astrophysics and Supercomputing, Swinburne University of Technology, Hawthorn, VIC 3122, Australia\\[0.1cm]
$^\dagger$adam.stevens@uwa.edu.au
}\\
\hrule

\author{Adam R.~H.~Stevens}

\section*{Abstract}

For astronomers to make a significant contribution to the reduction of climate change-inducing greenhouse gas emissions, we first must quantify our sources of emissions and review the most effective approaches for reducing them.  Here we estimate that Australian astronomers' total greenhouse gas emissions from their regular work activities are $\gtrsim$25\,ktCO$_2$-e/yr (equivalent kilotonnes of carbon dioxide per year).  This can be broken into $\sim$15\,ktCO$_2$-e/yr from supercomputer usage, $\sim$4.2\,ktCO$_2$-e/yr from flights (where individuals' flight emissions correlate with seniority), $>$3.3\,ktCO$_2$-e/yr from the operation of observatories, and $2.6\pm0.4$\,ktCO$_2$-e/yr from powering office buildings.  Split across faculty scientists, postdoctoral researchers, and PhD students, this averages to $\gtrsim$37\,tCO$_2$-e/yr per astronomer, over 40\% more than what the average Australian non-dependant emits \emph{in total}, equivalent to $\sim$5$\times$ the global average.  To combat these environmentally unsustainable practices, we suggest astronomers should strongly preference use of supercomputers, observatories, and office spaces that are predominantly powered by renewable energy sources. Where facilities that we currently use do not meet this requirement, their funders should be lobbied to invest in renewables, such as solar or wind farms.  Air travel should also be reduced wherever possible, replaced primarily by video conferencing, which should also promote inclusivity.

\begin{multicols}{2}

\section{Introduction}
\label{sec:intro}

Climate change is widely regarded as the biggest ongoing issue facing the planet's populous right now.  So much so, that over 11,000 scientists from 153 countries recently signed a paper warning of a global climate emergency\citep{ripple19}.
Humanity's continuing emission of greenhouse gases -- driven predominantly by the burning of fossil fuels as a source of energy\citep{fried19} -- has already led to a rise in the mean global surface temperature of $\sim$1$^\circ$C relative to pre-industrial levels\citep{ipcc1}.
For global heating to be limited to 1.5--2$^{\circ}$C per the \href{https://web.archive.org/web/20191113214239/https://unfccc.int/process/conferences/pastconferences/paris-climate-change-conference-november-2015/paris-agreement}{Paris Agreement} requires a decrease to effectively \emph{zero anthropogenic emissions} in the next few decades\citep{matthews08,allen18,forster18,rogelj18}. Even then, it is expected that there will be long-lasting (time-scales of $\gtrsim\!10^{3}$--$10^5$\,yr) or potentially permanent changes to the environment\citep{archer08,solomon09,depaolo15}, which will have (and are already having) widespread, significant impacts on many forms of life.
This has been discussed in the literature and media for decades, with a complete technical elaboration provided as part of the fifth Assessment Report from The United Nations' Intergovernmental Panel on Climate Change\citep{ipcc_ar5}.

As is the case for most (if not all) professions, there are many aspects to being an astronomer that currently result in the emission of greenhouse gases and, therefore, a direct contribution to climate change.  Broadly, these include direct emissions from flights, and indirect emissions from the electricity required to power supercomputers, observatories, and other facilities, in addition to emissions associated with their construction.  We are no less responsible for ensuring we reduce our emissions from these activities than anyone else in the world is for reducing their own sources of emissions.

To address methods for emissions reduction demands that one understands not only where their own sources of emissions come from, but also what their relative quantitative significance is.
Part of the purpose of this Perspective is to provide astronomers with a base level of quantitative information on their sources of emissions.
More than this though, it is imperative that acknowledgement of this leads to action that will result in a decrease in the community's emissions.
For to be aware of a problem but choose not to act is practically no different than to deny the problem's existence, especially when one is demonstrably contributing to said problem\citep{cohen01,walker19}.
We all have an ethical obligation here that must not be ignored.

Climate change action is particularly important for Australia-based astronomers (and Australians in general), as Australia's record of greenhouse gas emissions is particularly poor in the global context. Australia's total emissions (excluding international flights and shipping) for the year ending March 2019 were 538.9 million equivalent tonnes of CO$_2$ (MtCO$_2$-e)\citep{gov19}.
With a population of 25.287 million people at the end of the March 2019 quarter 
\href{https://web.archive.org/web/20191130225413/https://www.abs.gov.au/ausstats/abs@.nsf/mf/3101.0/}{according to the Australian Bureau of Statistics} 
-- of which 
\href{https://web.archive.org/web/20200123110724/https://www.abs.gov.au/ausstats/abs@.nsf/0/1CD2B1952AFC5E7ACA257298000F2E76?OpenDocument}{$18.7\%$ are dependants under the age of 15\,yr} --
the country's emissions rate equates to 26.2\,tCO$_2$-e/yr per non-dependant.
%(throughout, we use the common definition of `non-dependant' to mean `person aged 15+', focussing on emissions per non-dependant rather than per capita, as this provides a more meaningful comparison to adults of working age)
This is in stark contrast to the 2018 global average of $7.3\pm0.7$\,tCO$_2$-e/yr per non-dependant (based on total emissions from the Global Carbon Budget 2019\citep{fried19} and the global population from \href{https://www.worldometers.info/world-population/}{Worldometer}, taking half the range of the 2017 and 2019 values as the uncertainty on the latter)
and makes Australia one of the highest-emitting countries per person in the world.  Countries that have comparable per-capita emission rates to Australia include the United States and Canada\citep{ritchie19}.  Perhaps it is no coincidence then that members of the astronomical communities from these countries have written white papers on this same topic, which include several practical, sensible suggestions for mitigation strategies\citep{matzner19,williamson19}. This is clearly an issue that astronomers worldwide are cognisant of; the Canadian paper\citep{matzner19} was one of the 5 most widely discussed papers for its month of release, with members from 43 astronomy institutes up-voting it on \href{https://voxcharta.org/?s=1910.01272}{Voxcharta}.
In Australia, \href{https://web.archive.org/web/20200319035744/https://laureatebushfiresclimate.wordpress.com/}{an open letter} has been written to the federal government, highlighting the urgent need to reduce greenhouse gas emissions, which has been signed by over 80 Laureate Fellows -- the most senior and prestigious research fellows funded by the Australian Research Council -- including 6 astronomers.

In this Perspective, we take approximate stock of the greenhouse gas emissions for which Australian astronomers are responsible (Section \ref{sec:srcs}). We then present options for how these sources may (and should) be reduced, and discuss current initiatives along these lines (Section \ref{sec:solns}). Our focus on the Australian community is a practical one, as it reflects the fact that we -- the authors -- are all based in Australia.  Despite this, the underlying content and message of this paper should be relevant for the global astronomical community and other fields of science.

\section{Sources of astronomers' emissions}
\label{sec:srcs}

In this section, we provide an overview of the emissions that Australian astronomers are responsible for, from the sources of expected greatest significance, in no specific order.

\subsection{Flights}
\label{ssec:flights}

Relative to the general public, astronomers travel \emph{a lot}.  Reasons include, but are not limited to:~conferences, workshops, collaboration, seminars, observing runs, committee meetings, job interviews, and relocation.
This is not specific to astronomers though; academics in general are responsible for significant greenhouse gas emissions from flying.  One case study suggests that business-related flights from university employees contribute approximately two thirds of the emissions of campus operations\citep{wynes18}. Flights are often the greatest \emph{single} source of university emissions, with conference attendance accounting for approximately half of those flight emissions\citep{middleton19}.

Not only does \emph{all} international travel require flying thousands of kilometres from Australia, but due to the size and low population density of the country, domestic travel often does too. 
As a point of reference, we collate the approximate greenhouse gas emissions per passenger from direct flights between Australian capital cities in Table \ref{tab:qantas}, according to \href{https://www.qantasfutureplanet.com.au/}{Qantas}.  Based on the same carbon calculator, return trips from Australia to Europe or the Americas can comfortably exceed 3\,tCO$_2$-e per passenger.

In Australia, aviation was responsible for 22.02\,MtCO$_2$-e of emissions in 2016 alone (which includes 12.02\,MtCO$_2$-e from international flights)\citep{gov17}.  This suggests that aviation is responsible for $\sim$4\% of the country's total emissions (or close to 1\,tCO$_2$-e/yr per person on average).  
While this may sound like a small fraction, it is important  to recognise that \href{https://web.archive.org/web/20180111231455/http://www.roymorgan.com/findings/7084-sky-high-australians-air-travel-habits-201612091252}{about half the population will not fly at all in a given year, that most of them will only fly once in that year, and that the vast majority will do so for leisure, not business}. 
For the relatively few people who fly regularly, their personal fraction of emissions from air travel presumably must be \emph{much} higher than the nominal 4\%.  As we demonstrate below in this section, astronomers are among those people (at least, certainly in Australia).

\begin{table}[H]
\centering
\begin{tabular}{r | r r r r r r} \hline\hline
& ADL & BNE & CBR & HBA & MEL & PER \\ \hline
BNE & 340 &  &  &  &  &  \\
CBR & 240 & 306 &  & &  &  \\
HBA & 378 & 314 & 296 & & & \\
MEL & 134 & 288 & 122 & 144 &  &  \\
PER & 442 & 748 & 674 & 656 & 528 &  \\
SYD & 246 & 158 & 82 & 236 & 148 & 652  \\ \hline\hline
\end{tabular}
\caption{Typical emissions for (the most) direct return flights between Australian capital cities, according to \href{https://www.qantasfutureplanet.com.au/}{Qantas}.  Three-letter names are the official airport codes.  Units for emissions are kgCO$_2$-e per passenger (rounded to the nearest integer).}
\label{tab:qantas}
\end{table}

\subsubsection{CAS budget example}
\label{ssec:budget}

As an example of astronomy's disproportionately high flight emissions, consider Swinburne University of Technology's Centre for Astrophysics and Supercomputing (CAS). In 2017, approximately 80\% of CAS's travel budget was spent on flights:~$\sim$\$301k in total (including external funding contributions), with \$54k spent on 134 domestic round-trip flights, and the remaining \$247k spent on 133 international round trips (often including more than two flights). These flights covered the $\sim$80 full-time-equivalent (FTE) staff and students in CAS during 2017, meaning each person was responsible for $\sim$1.7 domestic and $\sim$1.7 international flights on average. A typical domestic return flight from Melbourne produces $\sim$230\,kgCO$_2$-e per passenger (taking a na\"{i}ve average of the values for MEL in Table \ref{tab:qantas}). Considering Los Angeles as a typical international destination, a return international flight produces $\gtrsim$3\,tCO$_2$-e per passenger (per \href{https://www.qantasfutureplanet.com.au/}{Qantas's calculator}). Therefore, the average astronomer in CAS was responsible for $\sim$5.4\,tCO$_2$-e in 2017 from flying alone, with 0.4 and 5.0\,tCO$_2$-e coming from domestic and international flights, respectively. As a rough guide to the average monetary carbon cost of flying, these figures imply $\sim$0.57\,kgCO$_2$-e per AUD for domestic flights and 1.6\,kgCO$_2$-e per AUD for international flights.  These figures are comparable to the case study of \citet{stohl08} at a different institute (and research field).

\subsubsection{ICRAR-UWA travel records}
\label{ssec:survey}

A further, more detailed example is available from the International Centre for Radio Astronomy Research -- University of Western Australia node (ICRAR-UWA).  Here, the complete travel records for the 2018 and 2019 calendar years were analysed.
Over this time, ICRAR-UWA used three different travel agencies.  All work-related travel captured by these agencies was accounted for, regardless of the funding source.  Two of those agencies gave direct emissions values for all bookings captured by their systems.  For the third agency, we still had access to all flights travelled, but had to calculate the emissions for each flight; for this, we used \href{https://www.qantasfutureplanet.com.au/}{Qantas's calculator}.

All emissions initially quoted did not differentiate between economy and business class flights.  Business class seats occupy roughly triple the area of economy seats (\href{https://www.qantas.com/au/en/qantas-experience/onboard/seat-maps.html}{this varies plane to plane}, and is often lower for domestic trips and higher for international trips, with \href{https://web.archive.org/web/20200117133313/https://www.theguardian.com/environment/blog/2010/feb/17/business-class-carbon-footprint}{one article} suggesting a factor of 3.5 is more common for the latter).  For the relatively few business class flights listed in the travel records, we multiplied their emissions by 3.  We emphasise that economy class is the norm for astronomers, and the vast majority of bookings in these records were indeed economy.

\begin{figure}[H]
\centering
\includegraphics[width=1.01\textwidth]{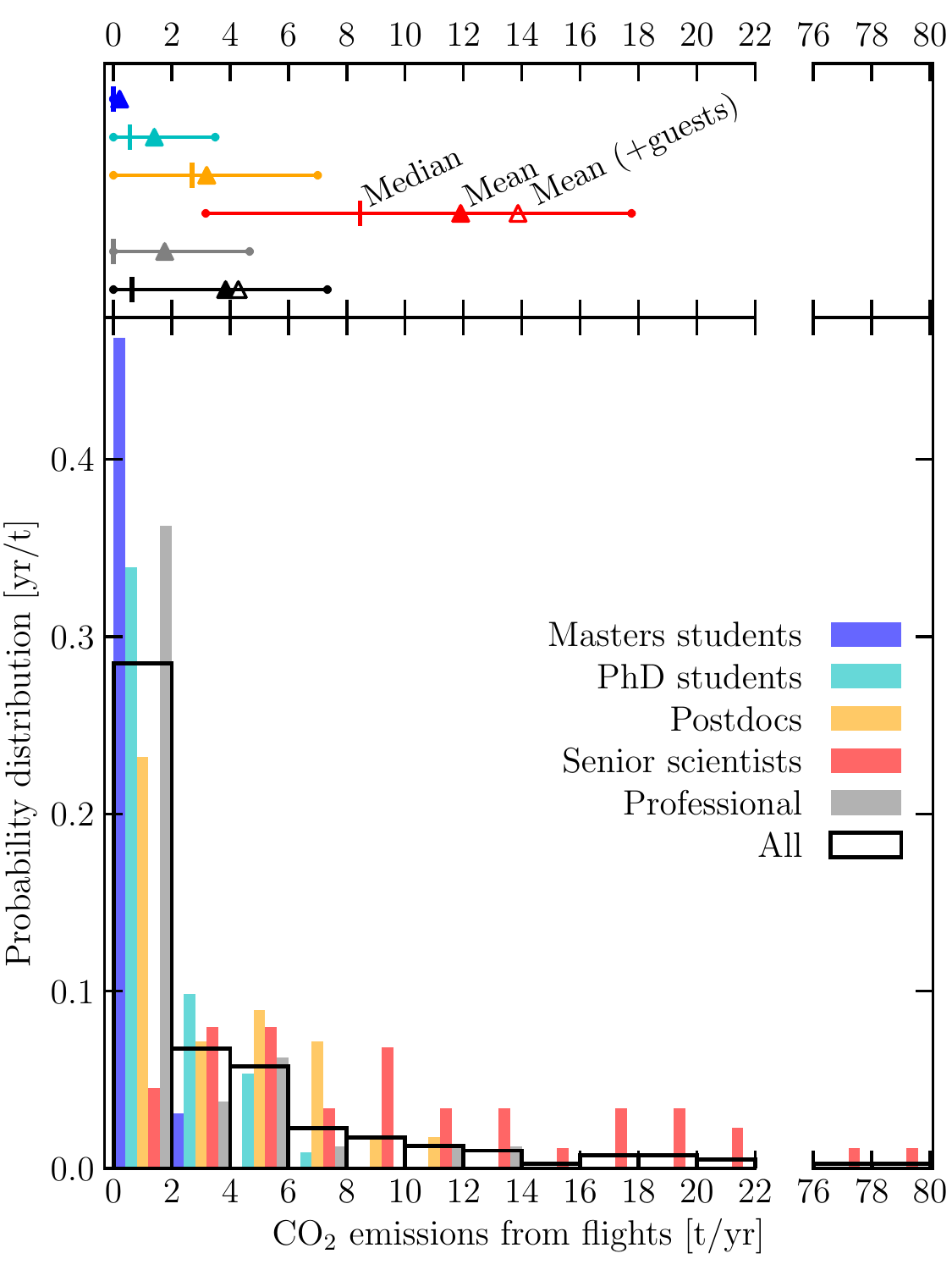}
\caption{Distributions of ICRAR-UWA staff's air travel CO$_2$-equivalent emissions.  Bottom panel: normalised histograms of individuals' annual emissions for 2018 and 2019 (2 entries per person) from each staff type in bins of width 2\,t/yr; bar thicknesses (except for the `all' category) have been artificially reduced to visually separate each distribution.  Top panel: mean of each distribution (closed triangles, where open triangles add the contribution from visitors), along with their medians (tall, vertical dashes), and 16$^{\rm th}$ \& 84$^{\rm th}$ percentiles (dots connected by horizontal bars).  These values are provided in Table \ref{tab:flights}.}
\label{fig:dists}
\end{figure}

In Table \ref{tab:flights} and Fig.~\ref{fig:dists}, we summarise the findings from ICRAR-UWA's travel records.  While these data have been anonymised, we present statistics for different levels of staff.  Where we refer to `senior scientists', we mean all research staff employed at Level C and above in the Australian university employment system, which are effectively all tenured or tenure-track positions, including senior fellows, associate professors, and full professors.  We broadly label all nominal research staff employed at Level A or B as a `postdoc', all of whom are on fixed-term contracts, which includes research associates and early-career fellows.  All remaining staff who are not students fall under the `professional' category.  This covers a diverse range of staff, including outreach, administration, computer scientists, and engineers.  Masters and PhD students are considered separately.  Other students are not explicitly accounted for (e.g.~Honours students, of which ICRAR-UWA has none).  

Unsurprisingly, flight frequency -- and thus flight emissions -- scales with seniority (as has been found in other studies\citep{wynes18}).  The average senior staff member emits close to 12 equivalent tonnes of CO$_2$ from flying each year (or roughly four return international trips, or three international + four domestic).  Granted, this mean (but not the percentiles) is pulled up by two outlier points in the distribution; removing the factor-of-3 assumption regarding emissions of business versus economy seats would reduce this mean to 9.5 tonnes. The average postdoc emits around a third that of an average senior staff member (roughly, one international and one domestic trip each year). The average PhD student emits less than half that of an average postdoc (2--3 domestic trips each year, or 1 international trip every two years).

In total, the flight emissions from ICRAR-UWA staff members over the two-year period was 768\,tCO$_2$-e.  A further 86\,tCO$_2$-e came from guest bookings, i.e.~travel booked by ICRAR-UWA staff for external visitors and collaborators.  It is important that these bookings are not ignored, because if the same study were conducted at those guests' home institutes, those flights likely would not be captured by their systems.  Likewise, there could well be other work-related flights that ICRAR-UWA staff members took over this period that were booked externally, thus not considered here.  Incorporating captured guest flights into our figures compensates for this. In all instances,  a senior member of staff was the host for the guests, so this reasonably should only contribute towards the figures for senior staff and totals.  We include a second column for means in Table \ref{tab:flights} that appropriately takes guest flights into account.

Remarkably, the average per-person emissions from flights of PhD students, postdocs, and senior scientists combined at ICRAR-UWA is exactly the same as the estimate for CAS (Section \ref{ssec:budget}), i.e.~5.4\,tCO$_2$-e/yr (excluding guests' flights).  After adding guest bookings, this average increases to just over 6\,tCO$_2$-e/yr.

Despite Perth's relative isolation (it is the second-most isolated major city globally, based on nearest-neighbour distance of cities with populations above 1 million), the travel budgets of research institutes in Perth do not necessarily exceed that of equivalent institutes elsewhere in Australia.  While a domestic trip for those living on the continent's east coast might mean lower emissions (see Tab-%
\noindent le \ref{tab:qantas}), this is likely counterbalanced by an increase in the \emph{number} of domestic trips.  International travel comes at a heavy carbon cost regardless of the Australian city of origin.

\begin{table*}[t!]
\centering
\begin{tabular}{l c c c c c c} 
\hline\hline
Staff type & Number & \multicolumn{5}{c}{CO$_2$ emissions per person [t/yr]}\\
& & $16^{\rm th}\,\%$ile & Median & $84^{\rm th}\,\%$ile & \multicolumn{2}{c}{Mean(s)}  \\ 
\hline
Masters students & 16 & 0.00 & 0.00 & 0.00 & 0.21(0) & \\
PhD students & 28 & 0.00 & 0.55 & 3.49 & 1.40(0) & \\
Postdocs & 14 & 0.00 & 2.70 & 7.00 & 3.16(2) & \\
Senior scientists & 22 & 3.16 & 8.45 & 17.77 & 11.91(6) & 13.87(6)\\
Professional & 20 & 0.00 & 0.00 & 4.67 & 1.76(1) & \\
\hline
Senior+Postdoc+PhD & 64 & 0.00 & 3.12 & 9.51 & 5.40(1) & 6.07(1)\\
All & 100 & 0.00 & 0.63 & 7.34 & 3.84(0) & 4.27(0)\\
\hline\hline
\end{tabular}
\caption{Summary of greenhouse gas emissions from ICRAR-UWA employees' work-related flights from 2018 and 2019.  Emissions are measured in equivalent tonnes of CO$_2$ per person per annum.  See Section \ref{ssec:survey} for a full description of each staff type.  The second-to-last row combines results from PhD students, postdocs, and senior scientists.  The final row does the same, with the additional inclusion of Masters students and professional staff.  The second mean column (last column) adds the contribution from guest flights that were captured by the ICRAR-UWA booking system.  Uncertainties on the last digit for the means (bracketed numbers) are calculated from jackknifing.}
\label{tab:flights}
\end{table*}

\subsubsection{National extrapolation}
\label{ssec:extrap}
Official figures submitted as part of the 2019/20 mid-term review of the Australian astronomy decadal plan suggest there are currently 365.2 FTE research staff nationwide, covering academic levels A--E, i.e.~junior postdocs through to full professors.  These figures \href{https://www.science.org.au/supporting-science/science-policy-and-analysis/decadal-plans-science/decadal-plan-australian/white}{will be made public} as part of the mid-term review process.  Consistent with our earlier definition, if we consider postdocs to hold temporary contracts and be employed at either academic level A or B, then postdocs account for 166.2 of those FTEs.  That leaves 199 `senior scientists', which we again consider as those at academic level C and above, and/or those with permanent employment.  5 additional FTEs fall outside the standard university employment levels, which we do not categorise here.  326.5 FTE astronomy PhD students are enrolled nationwide, as are 72 FTE Masters students.  An earlier figure from 2014 suggested 242 support staff were also employed across the country\citep{wg3.1}, which we equate to our `professional' category.

Combining these numbers with the means in Table \ref{tab:flights} (including the guest contribution to senior scientists) gives an estimate of the total national emissions from flights as 4190\,tCO$_2$-e/yr.

\subsection{Supercomputer usage}
\label{ssec:super}

As described in a recent white paper\citep{otoole19}, the estimated computing requirements of Australian astronomers is 400 million CPU core-hours (MCPUh) per annum, expected to rise to 500\,MCPUh/yr by 2025.  This is split across many computing facilities, including both domestic and international supercomputers.  Each has its own energy efficiency and is powered by different sources.  It is therefore non-trivial to translate this level of computer processing into a rate of CO$_2$-equivalent emissions.

The three most significant supercomputing centres for Australian astronomers are the National Computing Infrastructure (NCI) in the Australian Capital Territory (ACT), the Pawsey Supercomputing Centre in Western Australia (WA), and the \emph{OzSTAR} supercomputer in Victoria.
We contacted each of these to request official figures on the energy/emission requirements that would allow us to estimate astronomers' computing carbon footprint as accurately as possible.  Unfortunately, Pawsey was the only centre that responded with data.  We therefore extrapolate from these data to estimate the national computing emissions of Australian astronomers.

Figures provided to us privately by Pawsey show that the Centre consumed 10.94\,GWh~of electricity in the 2018/19 financial year, $<$100\,MWh of which came from their own solar panels.
While one of Pawsey's two solar inverters was down for much of this period, we can reasonably estimate that 99\% of the electricity powering Pawsey comes from the grid.
25\% of Pawsey's computing resources are allocated to astronomy through the dedicated \emph{Galaxy} supercomputer\citep{pawsey19}.  We can therefore estimate that Australian astronomers require 2.7\,GWh/yr of electricity for their Pawsey usage alone (this is likely a lower limit, as other machines at Pawsey -- e.g. \emph{Magnus} -- are used by astronomers too).
In south-west WA, electricity currently carries a carbon cost of 0.75\,kgCO$_2$-e/kWh (we account for both `scope 2' and `scope 3' emissions when considering mains power consumption throughout this paper; in principle, this includes the emissions associated with extraction and burning of the fuel used to produce the electricity, as well as losses in transmission)\citep{gov18b}.  
2.7\,GWh/yr at Pawsey therefore translates to 2.0\,ktCO$_2$-e/yr.  
51.1\,MCPUh were consumed on \emph{Galaxy} for radio astronomy during the 2018/19 financial year\citep{pawsey19}, implying a carbon cost of $\sim$40\,tCO$_2$-e/MCPUh.

Given the above, we estimate the net power required to run code on a supercomputer that includes all overheads and cooling to be $\sim$53\,W/core.  In theory, this value could actually be higher for many facilities, as \href{https://web.archive.org/web/20200310054946/https://pawsey.org.au/groundwater-cooling-system/}{Pawsey uses a groundwater cooling system} that should reduce the energy requirements of cooling.
Nevertheless, if we assume that 53\,W/core is typical for most supercomputers, then we need only consider where other commonly used facilities are, and the emissions per kWh there.  In Victoria and the ACT, electricity emissions are 1.17 and 0.92\,kgCO$_2$-e/kWh, respectively\citep{gov18b}.  Despite these being official numbers from the Australian Government, we highlight a significant caveat regarding the emissions from electricity use in the ACT below.  For now, we take those numbers at face value.  Assuming a ratio of 3:2:1 for NCI:Pawsey:\emph{OzSTAR} (ACT:WA:Victoria) usage in astronomy (a difficult ratio to gauge with publicly available information), this gives an average of 0.905\,kgCO$_2$-e/kWh or $\sim$48\,tCO$_2$-e/MCPUh.  

It is important to note that only $\sim$60\% of Australian astronomers' supercomputer usage is from domestic facilities\citep{otoole19}.  \href{https://web.archive.org/web/20200205150416/http://www.compareyourcountry.org/climate-policies?cr=oecd&lg=en&page=2}{The average emissions per kWh for countries in the OECD (Organisation for Economic Co-operation and Development) is roughly half that of Australia's.} Accounting for this -- assuming it reflects where the offshore supercomputers that Australian astronomers use are -- reduces the average emissions for Australian astronomers' supercomputing time to $\sim$38\,tCO$_2$-e/MCPUh.  

With all of this in mind, we estimate that the total emissions from Australian astronomers' supercomputer usage is $\sim$15\,ktCO$_2$-e/yr.  This is nearly \emph{quadruple} the value from flights (Section \ref{ssec:extrap}).  Dividing across all senior staff, postdocs, and PhD students gives a mean supercomputing carbon footprint of $\sim$22\,tCO$_2$-e/yr per researcher.  Note that we have implicitly assumed that cores on local clusters in Australia carry the same power and carbon requirements as cores on supercomputers; the 400\,MCPUh/yr figure should include the use of local clusters.  Similar to flights, we expect that many of the people being averaged over will require relatively negligible computing time, and thus the mean emissions per researcher will be much less than the actual emissions of the researchers who have a heavy reliance on high-performance computing.

While it is difficult for us to quote an uncertainty on this number within a specified confidence interval, we can take 28\,tCO$_2$-e/yr per astronomer as a fair upper limit (the figure we would have derived had we not accounted for the lower overseas emissions for electricity).  Because of the significant production of renewable energy that the ACT is responsible for\citep{cass19}, one can argue that emissions from NCI should be treated as zero.  Taking that argument, while maintaining the assumption that 30\% of Australian astronomers' computing is done in the ACT, would lead a value of $\sim$14\,tCO$_2$-e/yr per researcher.  This provides a reasonable estimate of a lower bound.

\subsection{Observatories and telescopes}
\label{ssec:observatories}
Another potentially significant source of emissions is the operation of observatories and telescopes.  We sought information from several observatories regularly used by Australian astronomers regarding their emissions from operations (e.g.~power consumption).  While the information provided to us is not a complete accounting of all relevant domestic and international observatories (not all places we contacted supplied data), we can place a meaningful lower limit on the total electricity and emissions requirements for Australian astronomers to conduct observations.

In private communication, the \href{https://www.atnf.csiro.au/}{Australia Telescope National Facility} (ATNF, part of CSIRO) provided us with the electricity consumption of all observatories they operate over a one-year timeframe.  The sites considered include the Australia Telescope Compact Array (ATCA), the Parkes Observatory, the Mopra telescope, and the Murchison Radio-astronomy Observatory (MRO).  ATCA, Parkes, and Mopra all use mains power (with back-up diesel generators) and are all situated in New South Wales (NSW).  Those three sites consumed a combined total of 3760\,MWh of electricity over the year ending 29 February 2020, including all the telescopes, buildings, and integral facilities onsite.  ATCA accounts for 1920\,MWh, with $\sim$70\% of its observing time allocated to Australia-based `Principal Investigators' (PIs) in 2018\citep{atnf19}.  Parkes accounts for 1550\,MWh, with $\sim$55\% allocated to Australian PIs in 2018\citep{atnf19}.  The remaining 290\,MWh covers Mopra, although it is harder to obtain a fraction of time spent by Australian PIs on this telescope.  An earlier report from 2015\citep{atnf15} shows 8 programmes were run on Mopra on the year prior, with 3/8 of the first-name observers identified as belonging to Australian institutions.  Given the carbon cost of 0.92\,tCO$_2$-e/MWh for mains power in NSW\citep{gov18b}, the combined operation of ATCA, Parkes, and Mopra produces $\sim$3.5\,ktCO$_2$-e of emissions per year, with a contribution based on Australian astronomers' usage of $2.2\pm0.1$\,ktCO$_2$-e/yr.

The MRO hosts both the Murchison Widefield Array (MWA) and the Australian Square Kilometre Array Pathfinder (ASKAP).
The isolation of the MRO in Western Australia means it is not connected to mains power.  Instead it is powered by a combination of onsite solar photovoltaics and diesel.  Once operating at maximum capacity, the solar array is expected to cover $>$40\% of the site's electricity needs.  As of yet, it has not reached this capacity.  Over the 2018/19 financial year, the MWA and ASKAP consumed a total of 4110\,MWh of electricity:~$\sim$3360\,MWh for ASKAP, $\sim$520\,MWh for the MWA, and 230\,MWh from transmission losses.  600\,MWh of this came from solar energy, and the rest from diesel.  An additional $\sim$200\,MWh was consumed at the Boolardy accommodation facility, with roughly a third of this estimated to come from solar, and the rest diesel.  Based on figures from the Australian Government\citep{gov18b}, the carbon cost of burning diesel for energy is 266\,kgCO$_2$-e/MWh (this covers `scope 1' and `scope 3' emissions, i.e.~the onsite emissions from the burning of diesel and an approximate consideration of indirect emissions associated with its production and transport; the latter is likely an underestimate in the case of the MRO).  This implies the MRO currently produces greenhouse gas emissions at a rate of $\sim$0.95\,ktCO$_2$-e/yr.  Based on the facts that \href{http://www.mwatelescope.org/data/observing}{87.5\% of MWA observing time was led by Australia-based PIs in 2019} and $\sim$100\% of current ASKAP operations are Australia-led, we estimate Australian astronomers' contribution to MRO emissions as $\sim$0.93\,ktCO$_2$-e/yr.  Because the MRO is one of the sites for the Square Kilometre Array (SKA), its power consumption is expected to notably increase with time as SKA operations ramp up.  An increased fraction in dedicated solar power will help offset any rise in the site's emissions though.

The W.~M.~Keck Observatory in Hawai`i provided us with an estimate of their CO$_2$ emissions from on-site electricity and vehicle use. The latter is only a minor contributor. No flights to or from the Observatory were included (flights to the Observatory made by Australian astronomers have already been accounted for in Section \ref{ssec:flights}). The total CO$_2$ emissions reported to us were reduced pro-rata with Australia's current official proportion of Keck observing time of 10 observing nights. Noting that Keck operates two near-identical telescopes, there are 730 possible observing nights per (non-leap) year. Given this, an initial estimate of Australia's share of Keck's CO$_2$ emissions is 35\,t/yr.  Evidently, this is very small compared to emissions from Australia's use of its own domestic facilities.  The Australian astronomical community has had access to up to 40 nights per year on Keck in the past, but even the emissions from that would be almost negligible compared to the sum of ATNF observatories.

We note that the European Southern Observatory (ESO) has already commissioned a study of the emissions of its sites, but the results were pending at the time of writing this article.  Should these be made publicly available, this could prove a useful resource from 2020 onwards.  Now a Strategic Partner of ESO, Australia's emissions contribution to the use of those observatories should be taken into account for completeness.

With the information we have, we can confidently place a lower limit on the observatory-based emissions of Australian astronomers of 3.3\,ktCO$_2$-e/yr.  Contributions from the Siding Spring Observatory [which hosts the Anglo-Australian Telescope, the ANU (Australian National University) 2.3\,m telescope, the SkyMapper Telescope, and the UK Schmidt Telescope] and ESO facilities are the most significant exclusions from this estimate.  Any involvement that Australian astronomers have in space telescopes has also not been taken into account here.

\subsection{Campus operations}
\label{ssec:campus}
Office spaces and their machinery also contribute to work-related carbon emissions.  While a specific analysis of all the buildings that house astronomy departments in Australia is left for future investigation, we can again use ICRAR-UWA as an example and extrapolate.  ICRAR-UWA lies in the Ken \& Julie Michael Building at UWA.  Figures provided to us by UWA suggest that powering the entire building produces 618,772\,kgCO$_2$-e/yr.  ICRAR occupies 48\% of the building's floor area, implying the Centre is responsible for 297\,tCO$_2$-e/yr.  Given that 100 people have a desk at ICRAR-UWA (see Table \ref{tab:flights}), this implies an average of $\sim$3\,tCO$_2$-e/yr per person for office building requirements.  Extrapolating this to the $\sim$1000 astronomers and support staff nationwide (Section \ref{ssec:extrap}) implies total emissions of $\sim$3\,ktCO$_2$-e/yr.

A caveat to the building power requirements of ICRAR-UWA is that this includes powering the \emph{Hyades} computing cluster.  In principle, the emissions from the use of local clusters have already been accounted for in Section \ref{ssec:super}.  However, the entire \emph{Hyades} system only has 92 cores, meaning it must account for less than 0.8\,MCPUh/yr. Recognising that ICRAR-UWA makes up $\sim$10\% of the national community, any potential `double counting' of computing requirements must be less than 2\% of the total from Section \ref{ssec:super} (i.e.$<$0.3\,ktCO$_2$-e/yr).  In reality, local clusters like \emph{Hyades} almost never operate near their full capacity.

An additional caveat is some of the support staff who are based at observatories might already have their office requirements covered in Section \ref{ssec:observatories}.  It might be more appropriate to only extrapolate the per-person office power requirements to $\sim$800 people.  With both caveats, the true office-based emissions of Australian astronomers might be as low as $\sim$2.2\,ktCO$_2$-e/yr.

\subsection{Summary of emissions}
\label{ssec:summary}
Our findings are that the largest contributor to Australian astronomers' emissions is supercomputing.  At $\sim$15\,ktCO$_2$-e/yr for the national community (Section \ref{ssec:super}), this is more than all other sources of work-related emissions combined. This figure is primarily an extrapolation from power usage data we received from a single supercomputing facility (Pawsey).  There are many sources of uncertainty contributing to this figure that we have not quantified precisely.  With differing assumptions about how emissions from the ACT and non-Australian supercomputers are accounted for, this value could actually be as low as 9.5 or as high as 19\,ktCO$_2$-e/yr.

Despite sometimes garnering the most attention in conversation, flights rank a distant second (at best), totalling $\sim$4.2\,ktCO$_2$-e/yr (Section \ref{ssec:flights}).  This figure is based largely on an extrapolation of one institute (ICRAR-UWA), but it is entirely consistent with totals from a second institute (CAS).  The formal uncertainty carried through from the jackknifing uncertainties given in Table \ref{tab:flights} is effectively negligible, but it does not sufficiently account for potential variation across institutes in the country.  At a precision of one significant figure, we can fairly confidently say the value is near to 4\,ktCO$_2$-e/yr, assuming the values for emissions provided by travel agencies and airlines do not carry systematic uncertainties greater than $\sim$10\% (which we do not know).
Based on this, the uncertainty on our figure for flight emissions should be of order a few hundred tCO$_2$-e/yr.
However, we have not explicitly accounted for the altitude of aeroplane emissions, which is particularly problematic due to the production of contrails\citep{radel08,from12}.  In essence, the effective radiative forcing from aeroplane emissions at altitude \href{https://web.archive.org/web/20200319063138/https://www.eesi.org/papers/view/fact-sheet-the-growth-in-greenhouse-gas-emissions-from-commercial-aviation}{could be several times that of their nominal CO$_2$ emissions}.  This systematic is likely our greatest source of uncertainty.

The powering of observatories ranks third in emissions at $>$3.3\,ktCO$_2$-e/yr (Section \ref{ssec:observatories}).  This is based on the total power requirements of ATCA, the MRO, the Parkes Observatory, and Mopra -- accounting for the fraction of Australian PI time on these instruments -- with the additional but small contribution from Australia's time on Keck.
There are many other observatories that Australian astronomers use, and thus we can only provide a lower limit here.  In reality, the emissions from observatory operations could well exceed that of astronomers' flights.

Finally, emissions associated with powering astronomers' office buildings are approximated to be 2.2--3.0\,ktCO$_2$-e/yr nationwide (Section \ref{ssec:campus}).  Again, this is based on an extrapolation from ICRAR-UWA, and thus we may have underestimated the true uncertainty.

A visual summary of these four sources of emissions and their estimated uncertainties is provided in Fig.~\ref{fig:bar}. Summed together, the Australian astronomy industry is responsible for emitting $\gtrsim$25\,ktCO$_2$-e of greenhouse gases per year.  Dividing this across the combined 691.7 FTE of senior scientists, postdoctoral researchers, and PhD students implies an average of $\gtrsim$37\,tCO$_2$-e/yr per astronomer.  This means the work-based emissions of the average Australian astronomer exceed the \emph{combined work+life} emissions of the average non-dependant living in Australia by $>$40\%.
Globally, this is $\sim$5$\times$ the average work+life emissions per non-dependant.
Hypothetically, if half of all emissions were associated with people's work (and the other half their lifestyles), it would follow that an Australian astronomer's job is $\sim$3$\times$ as carbon-intensive as the average job in Australia, and $\sim$10$\times$ that of the average job globally.
While there are surely plenty of examples of other jobs that are equally or more carbon-intensive, no such comparison absolves anyone of responsibility.

\section{Solutions and discussion}
\label{sec:solns}

To contribute to the mitigation of unsustainable climate change, the astronomy community must focus on reducing the high rate of emissions found in the previous section.
In this section, we outline potential strategies to achieve this goal.

\subsection{Reduce flying}

Advances in aeroplane technology have helped to reduce the average emissions per passenger per unit distance in recent years.  From 2012 to 2016, the Australian aviation industry saw a 6.8\% increase in fuel efficiency\citep{gov17}.  This was, however, counteracted by a 16.8\% increase in fuel consumption due to a continual rise in airline traffic\citep{gov17}.  While there exist prospects for greater increases in fuel efficiency in future, this is not a domain that astronomers are likely to influence or accelerate.  The only practical action that ast-
\begin{figure}[H]
\centering
\includegraphics[width=0.95\textwidth]{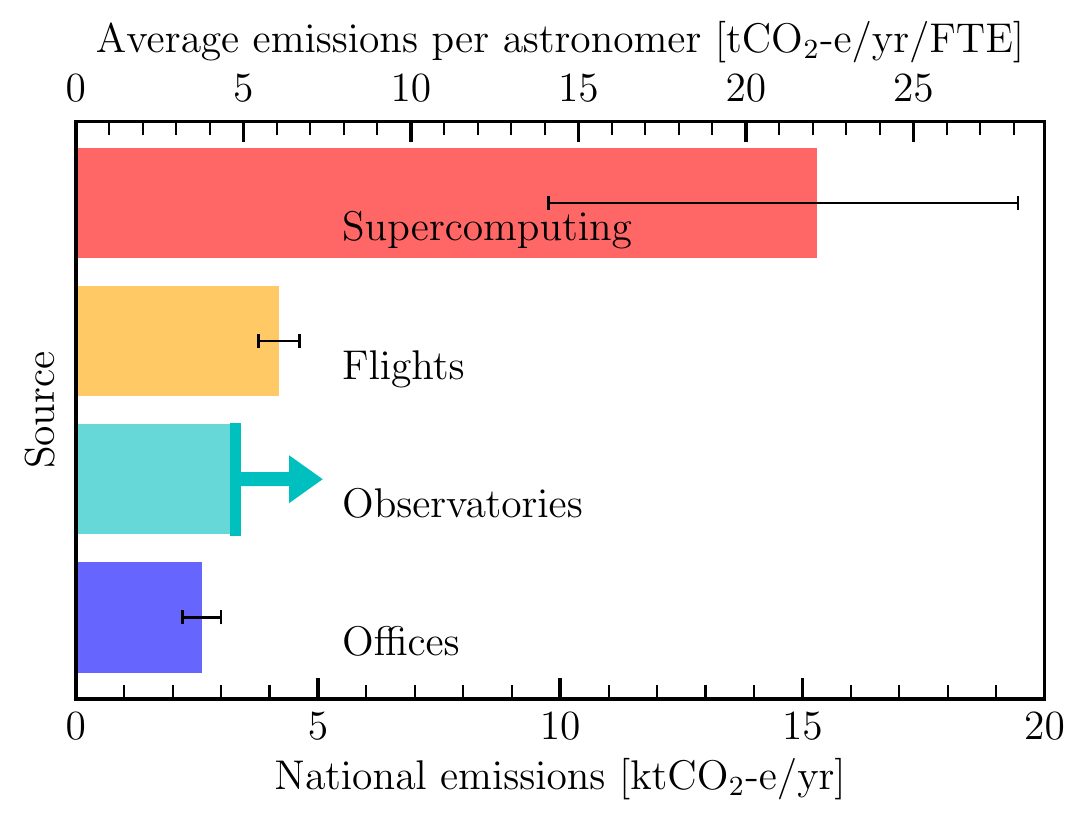}
\caption{Breakdown of the four sources of Australian astronomers' emissions considered in this work.  Error bars provide an estimate of our uncertainties, but should not be interpreted as formal confidence intervals.  The value for observatories is a lower limit.  `Per astronomer' refers to the 691.7 FTE including PhD students, postdocs, and senior researchers.  Values are summarised in Section \ref{ssec:summary}.}
\label{fig:bar}
\end{figure}
\noindent ronomers can take to reduce their flight emissions is to fly less.

The challenge then becomes: without flying, how do we achieve all the same things that up until now have involved air travel?  While potentially a confronting question for astronomers used to a high frequency of travel, events in 2020 have already demonstrated that this need not be a daunting challenge any longer.

At the time of finalising this paper, the outbreak of Coronavirus Disease 2019 (COVID--19) \href{https://web.archive.org/web/20200321223217/https://www.who.int/dg/speeches/detail/who-director-general-s-opening-remarks-at-the-media-briefing-on-COVID--19---11-march-2020}{was declared a pandemic}.  This has forced many people to cancel travel plans and work remotely.  Already, this is precipitating cultural change among astronomers, whereby online meetings have become commonplace.  
There exists an opportunity here to learn from this situation, enabling the global astronomical community to carry forward a low-travel, technology-focussed approach to communication and collaboration.
To achieve this, we need to be deliberate about not defaulting back to our previous travel habits.  The suggestions below consider what we can do (and/or should continue to do) once global restrictions on movement are lifted.

Perhaps the simplest and most obvious occasions where flying can be eliminated (or remain absent) are short meetings (1 or 2 days long) with a small number of people ($\lesssim$10), including, for example, time allocation committees and executive committees (\citet{wynes18} call for similar action).  
In principle, these could easily (continue to) be done via readily available video conferencing software.  
Despite the often quoted, yet anecdotal benefits of people's physical presence at meetings, the justification to fly thousands of kilometres for the sake of a short discussion is tenuous in the era of climate change action. 

An additional avenue by which our flight load can remain low is to (further) conduct observations remotely, rather than travelling to observatories. This practice is already being increasingly adopted globally. 
In Australia, this has been facilitated by the automation of facilities such as the ANU 2.3\,m telescope and Parkes radio telescope. Remote observing stations for the Anglo-Australian Telescope that are in several locations also help in this capacity. By having access to remote observing facilities at each of the major astronomy hubs in the country, not only can flights be reduced, but so too can accompanying financial costs.
For larger-scale, international facilities, the Keck remote observing room at Swinburne is open to the Australian astronomical community, and provides an alternative to international travel, even if it means domestic travel.  Observations conducted via ESO can be done in `service mode' or `designated visitor mode', nullifying the need to travel to ESO sites.

Inevitably, it seems conferences must move to a space where virtual attendance is also the norm.  Other research fields acknowledge this and have already started experimenting with online conferences (prior to COVID--19-driven social distancing)\citep{abbott20}.
To enable this, we must ensure conferences and meetings have adequate video conferencing systems available.  This could mean investing in either hardware and/or software to meet the requirements of running said meetings smoothly.
As a proactive example, ASTRO\,3D (the ARC Centre of Excellence for All Sky Astrophysics in 3 Dimensions) is currently considering whether the development of software beyond the capabilities of that regularly used by academics is warranted and worth funding.
Members of conference organising committees should not only have a plan for how they will make remote attendance possible, but also promote and/or advertise this as an option.
Indeed, several major astronomy meetings in 2020 will be run (or have already been run) entirely remotely because of COVID--19, including those of the European Astronomical Society and American Astronomical Society, each of which typically attracts of order 1000 participants.
Given that the logistical challenge of running a major conference online far exceeds that of conducting observations or a committee meeting, we should treat these conferences as opportunities to experiment, paying close attention to the aspects that work (i.e.~lead to a successful meeting, comparable to our experience of in-person conferences) and those that do not. 

An emphasis on virtual meetings has the added benefit of increased inclusivity.  Removing the need to travel enables those who are limited in their opportunity to travel (be it because of finances, health, carer responsibilities, or other reasons) 
to more readily participate.  Even for those without stringent limitations, a reduction in travel alleviates limitations on people's time, and thus increases participation opportunity for everyone.

We suggest that those wishing to travel should have to justify to their travel approvers (i) why alternatives to travel are unsatisfactory, and (ii) why their proposed trip is worthy of contributing to climate change.  If travel is approved, travellers should take careful note of flight options, as the route, airline, and aeroplane model can all influence the emissions per passenger of the journey.  Fewer flights with lower emissions should be preferenced over monetary savings per journey.

\subsubsection{On the ECR argument}
\label{ssec:ecr}
One reason often cited against flight reduction is that it might harm the careers of early-career researchers (ECRs) and late-stage PhD students.  After all, the astronomy job market is incredibly competitive, and the majority of astronomy PhDs will not find permanent positions in the field\citep{forbes08,metcalfe08,cooray15}.  A lack of exposure might therefore disadvantage job applicants, thereby becoming one of the many who ``don't make it'', despite being more than capable.  There are several problems with this argument.

For one, it is entirely anecdotal.  To our knowledge, there has not been a systematic study of the career pathways of astronomy PhDs and whether their frequency of flying in the early stages of their career had any effect on either their decision to stay in the field, or their ability to progress had they chosen to stay.  One could speculate that a minimal amount of international exposure might be necessary to get one's foot in the door, but the job-hiring and grant-winning processes are stochastic.  One could therefore equally speculate that, at some point, the probability of an application being successful as a function of the candidate's exposure might saturate.

The argument also encourages escalation.  Competitive people will always look for a way to stand out.  If we tell our students and ECRs that they will not stand out if they do not fly to speak at conferences and the like, then not only will they all fly, but the most competitive ones will find an additional means of outdoing their peers (which might mean flying even more).  Instead, we should focus on \emph{deescalating} the situation.  If it is globally mandated that flying should be minimised, then no ECR will be at any disadvantage to their peers by flying less, because everyone will be doing it.  In principle, this should have the added positive effect of alleviating some (but certainly not all) anxiety surrounding the overly competitive nature of astronomy:~one of the frequently cited reasons why people choose to leave the field\citep{arka18,karo18}.

It would help to build a culture where \emph{values} like environmental sustainability are not only supported, but are encouraged and factored into the job-hiring process (for a related discussion, see the paper by \citet{walkowicz18}).  Senior members of the community hold the greatest power in effecting this culture change.  They also have the greatest responsibility to reduce flight emissions, based on the numbers in Table \ref{tab:flights}, and have the least risk in doing so, given that their employment is ongoing.  While we encourage that ECRs should reduce their flying, the onus is not necessarily on ECRs in the first instance.

\subsubsection{On carbon offsetting}
Carbon offsetting is often cited as a method by which one can reduce their \emph{net} carbon footprint, be it from flying or other sources.  In essence, the idea is that by giving money to a scheme that will reduce emissions elsewhere or, ideally, help to remove greenhouse gases from the atmosphere, one offsets the emissions they are personally responsible for.  While not devoid of merit, both the principle and practice of carbon offsetting has been widely questioned.  Some critics, for example, have likened it more to purchasing absolution of guilt than having a tangible impact on greenhouse gas concentrations in the atmosphere\citep{hansen11,hodgkinson15}.

There are a wide range of offsetting schemes that exist.  It is often assumed that offsetting means planting trees, but this is rarely the case. In Australia, airline offsets tend to fund land conservation or fire abatement (see \href{https://web.archive.org/web/20190924072833/https://www.qantasfutureplanet.com.au/}{Qantas} and \href{https://web.archive.org/web/20191229121915/https://travel.virginaustralia.com/au/blog/celebrating-10-years-our-carbon-offset-program}{Virgin}, the country's biggest carriers).  While these are worthy causes for investment, their being funded simply \emph{prevents potential future} emissions (or prevents the reduction of the land's ability to sequester carbon from the atmosphere), and does nothing to remove the greenhouse gases added to the atmosphere from aeroplanes.  Even if all offsets were hypothetically funding reforestation, this would not solve climate change in a world where we continue to fly.  The solution to reducing the concentration of CO$_2$ in the atmosphere (and the ocean) requires \emph{both} reforestation \emph{and} emission reductions\citep{hansen11}.  

That said, provided those paying for offset schemes understand that it is not itself a solution, it is better to offset than not.  Of course, this does not have to be limited to air travel; if we are to offset our flights, we should also offset our power consumption (and other activities), especially that required for supercomputers, at least in the interim.  

It is important to choose and investigate an offset scheme carefully; it does not have to be affiliated with an airline.  Each astronomy department should consider a local scheme with tangible benefits to the environment, and ensure a fraction of their budget (travel or otherwise) is allocated for that scheme.

\subsection{Renewable energy sources}
\label{ssec:renewables}

Technology already exists for reducing our carbon footprint from supercomputer usage and other highly electricity-demanding operations.  It all comes down to what generates the energy.  Much of Australia's power comes from coal burning and other greenhouse gas-emitting sources\citep{gov18b}.  This is despite the fact that it has been known for years that it is feasible for Australia to be powered entirely by renewables\citep{elliston14,elliston16}, contrary to the narrative repeated by some of our politicians and other sceptics\citep{diesendorf18}. Realistically, it will take time for the country to continue its transition to renewables (as it will for the world).  We should, therefore, take action ourselves to ensure our electricity-demanding operations are powered by the greatest fraction of renewables possible.  

This means we should carefully choose the supercomputers we use, strongly favouring those certified as being powered predominantly by renewables.  Concurrently, we should be lobbying and/or helping the facilities we currently use to establish their own renewable energy sources.  An obvious first step would be to install solar panels at the facilities where they are not already present.
Some efforts in this direction have already been made.  For example, \href{https://web.archive.org/web/20190605045141/https://www.anu.edu.au/news/all-news/anu-invented-solar-tech-helps-university-reduce-emissions}{the roof of NCI holds 600 `sliver-cell' solar panels}, generating $\sim$93.5\,kW of carbon-free electricity.  The total power consumption of facilities like NCI is \emph{much} greater than this though.
Dedicated renewable energy farms that cover a much larger area than a building's roof (realistically, off-campus) are ultimately needed.
As mentioned in Section \ref{ssec:super}, Pawsey currently covers $\sim$1\% of its power with onsite solar photovoltaics, and is investigating options to increase their renewables' fraction in the foreseeable future.

As alluded to in Section \ref{ssec:super}, the ACT as a whole is now responsible for generating more renewable energy than the energy it consumes\citep{cass19}.  This power is not exclusively consumed in the Territory though; rather, it goes into the grid shared with NSW.  \href{https://web.archive.org/web/20191130225413/https://www.abs.gov.au/ausstats/abs@.nsf/mf/3101.0/}{For reference}, the ACT accounts for less than 2\% of the country's population, and is a factor of $\sim$19 less populous than NSW. One could argue that the operations of NCI should be considered carbon-neutral because its power consumption has (presumably) been accounted for in the ACT's renewables generation.  Equally though, any power drawn from the NSW+ACT grid increases the demand, and the supply that meets this is ultimately still backed by emissions-heavy power sources.
That is to say, if the operations of NCI were to cease (or reduce), there would be a measurable reduction in emissions.
As such, our default stance has not been to treat astronomers' usage of NCI as carbon-neutral (evidently, \href{https://web.archive.org/web/20190410090834/https://services.anu.edu.au/campus-environment/sustainability-environment/sustainability/energy}{the ANU does not treat NCI operations as carbon-neutral either}).
Nevertheless, initiatives to invest in renewables are precisely what we should be supporting.  By extension, it seems favourable to support supercomputing facilities that reside in areas whose local governments are of this philosophy.
Per Section \ref{ssec:super}, were we to assume that NCI is carbon-neutral, our figure for the total emissions of Australian astronomers would drop by nearly 6\,ktCO$_2$-e/yr.

Observatories should also be powered by renewables, which several observatories have already recognised.  As mentioned in Section \ref{ssec:observatories}, the MRO has a dedicated hybrid solar--diesel power station, \href{https://web.archive.org/web/20190306003156/http://www.mwdc.wa.gov.au/6/377985.ashx}{with the potential to supply the site with up to 50\% renewables} (although currently sits at closer to 15\%).  ESO's \href{https://www.eso.org/public/about-eso/green/}{La Silla Observatory has a dedicated solar farm} on site too, as \href{https://www.eso.org/public/announcements/ann19057/?lang}{will ESO's Extremely Large Telescope}.   

Many universities in Australia have set targets for approaching `carbon neutrality'.  \href{http://www.news.uwa.edu.au/2019091911618/energy-carbon-neutral-uwa-your-campus-your-future}{UWA aims to have its electricity requirements fully covered by renewables by 2025}, with plans to further offset other sources of emissions by 2030.  With a slightly more accelerated timeline, \href{https://web.archive.org/web/20191007035744/https://sustainablecampus.unimelb.edu.au/key-areas/energy}{The University of Melbourne aims to be energy carbon-neutral by 2021, and fully neutral by 2030}.  The 2030 goal is also shared by \href{https://web.archive.org/web/20191115124627/https://www.monash.edu/net-zero-initiative}{Monash University}.  Swinburne University \href{https://www.swinburne.edu.au/news/latest-news/2020/01/swinburne-supports-climate-emergency.php}{plans to procure 100\% renewables by mid 2020, and be carbon-neutral by 2025}; perhaps most significantly for astronomers, this will include covering the energy requirements of the \emph{OzSTAR} supercomputer.  The University of Queensland has its own off-site solar farm, which was planned \href{https://web.archive.org/web/20190403211612/https://www.uq.edu.au/news/article/2018/06/uq-set-world-standard-100-cent-renewable-energy}{to make the University energy carbon-neutral by 2020}, while the \href{https://web.archive.org/web/20200310085715/https://www.sustainability.unsw.edu.au/our-plan/climate-action}{University of New South Wales wants to purchase all its electricity from existing renewables} by 2020.  The \href{https://www.anu.edu.au/news/all-news/anu-council-says-we-must-act-on-climate-change}{latest announcement by the ANU} states an intent to become `net negative' in their emissions, although the time-scale to achieve this is yet unclear.
Whether the various initiatives of these universities pan out as planned remains to be seen.  We can all place pressure on our universities to ensure these policies are seen through or even accelerated where possible.

\subsection{Create incentives}
While the ethical and scientific arguments for significant action to reduce our contribution to climate change undoubtably have an impact on individuals, the lack of tangible action on this topic thus far suggests that action at an institutional/governing level is also necessary. 

To perhaps state the obvious, creating additional incentives to reduce carbon emissions should, in principle, help to reduce carbon emissions.
One option is to establish an award that departments set out to earn.  This could be based on purely having low emissions, or could more broadly encompass environmental sustainability.  The Astronomical Society of Australia (ASA) went through the same process for a different area of ethical importance several years ago: the Pleiades award for gender equity and diversity.  
The movement of promoting diversity and equity has resulted in focussed committees at most major institutions, \href{https://web.archive.org/web/20190929105542/https://www.industry.gov.au/news-media/science-news/australias-first-women-in-stem-ambassador}{an ambassador for women in STEM} (science, technology, engineering and mathematics), and numerous national programmes to tackle this problem, including the \href{https://www.sciencegenderequity.org.au/}{Science in Australia Gender Equity initiative}. We, as a community, should work towards a future where the same importance is placed on the planet we live on as the people that live on it, to make sure that our legacy is more than just academic. 
Given the success of the Pleiades and Athena SWAN (Scientific Women's Academic Network) awards\citep{kewley19}, a low-emissions award could be modelled directly on them.  The ASA is an ideal organisation to lead this because (i) it is a national body, and (ii) it exists in perpetuity, unlike other national entities (such as Centres of Excellence).

%Carbon-efficient research can also be incentivised by supercomputing and telescope time allocation committees.  An application's merit could partly be judged on a statement regarding the ratio of potential research impact to emissions produced for the proposed project.  Evidence for a plan to minimise a project's potential emissions should be favoured.  Track records of investigators' outputs from earlier low-carbon research could also be specifically taken into account.  Similar statements could be made compulsory for job applications too, relating back to the discussion in Section \ref{ssec:ecr}.

\subsection{Goal setting}
The Paris Agreement lays out goals for emissions reductions on several time-scales that will quantifiably limit the rise of the global mean temperature.  Loosely, the primary goals are to reduce emissions from 2018 rates by $>$50\% by 2030 (or an annual reduction of 7.6\% every year this decade), and 100\% by 2050\citep{rogelj18,unep19}.  This should keep global heating below 1.5$^\circ$C.

One option for the astronomy community is to follow the Paris Agreement percentage targets.  However, acknowledging that Australian astronomers' work-related emissions exceed that of the average adult's globally by an order of magnitude (Section \ref{ssec:summary}), our percentage goals should arguably be even bigger; practically, those who emit more have greater potential to reduce emissions, not just absolutely, but also fractionally.  A plan for all supercomputers, observatories, and offices that Australian astronomers rely on to be powered entirely by renewables would already see a $\sim$90\% reduction in the community's emissions.  It is not unfathomable that this could be achieved by 2030, especially as many Australian universities' carbon-neutral plans are already in motion (Section \ref{ssec:renewables}).  These will help cut up to 45\% of astronomers' emissions, based on the sum of those associated with offices, NCI, and \emph{OzSTAR} shown in Section \ref{sec:srcs} (this fraction will be notably less if ACT emissions were already to be treated as zero).  With active commitment moving forward, the community is well positioned to make a real contribution to limiting the effects of climate change.  Let's do it!

\section*{Data availability}
Travel records for ICRAR-UWA staff and students, and ATNF electricity data are private.  Queries about how the former were processed should be directed to A.R.H.S.  Similarly, flight records of CAS and data from Keck are private, but queries regarding these can be directed to M.T.M.  Power meter data from Pawsey are also private; requests for these data should be directed to Pawsey themselves, and we would encourage copying in P.J.E.  The demographics of Australian astronomers will be made \href{https://www.science.org.au/supporting-science/science-policy-and-analysis/decadal-plans-science/decadal-plan-australian/white}{publicly available online} with an accompanying white paper; none of the authors here are involved.

\section*{Ethics statement}
The use of ICRAR-UWA travel data (Section \ref{ssec:survey}) was approved for this study via an ethics exemption from the Human Ethics office at The University of Western Australia.  Although the data are anonymised in this study, specific informed consent was obtained from anyone deemed potentially identifiable.  Appropriate permission from the Centre for Astrophysics and Supercomputing at Swinburne University of Technology was obtained for the use of information presented in Section \ref{ssec:budget}.

\section*{Acknowledgements}
S.B., P.J.E., and M.T.M. are all supported through the Australian Research Council, respectively through project numbers DP180103740, CE170100013, and FT180100194.  

This work is an update from a white paper that was commissioned as part of the \href{https://www.science.org.au/supporting-science/science-policy-and-analysis/decadal-plans-science/decadal-plan-australian/white}{mid-term review of the 2016--2025 decadal plan for astronomy in Australia}.  
A.R.H.S.~thanks L.~Staveley-Smith and C.~Trott in their capacity as members of the mid-term review panel for their encouragement to write the paper.   

A.R.H.S.~and S.B.~thank R.~Sharma for extended help in obtaining data pertinent to calculating ICRAR-UWA's emissions.  P.J.E.~thanks B.~Evans and the Pawsey Supercomputing Centre for generously supplying data used in this work.  A.R.H.S.~thanks P.~Mirtschin and P.~Edwards at CSIRO for volunteering and providing information on ATNF facilities.

Figs~\ref{fig:dists} and \ref{fig:bar} were built with the {\sc matplotlib}\citep{hunter07} package for {\sc python}.

All authors thank the referees for their considered comments, which helped to improve this paper.

\section*{Author contributions}
A.R.H.S.~led the writing, analysis, and design of this work, but all authors made important contributions to all aspects of this paper.  
S.B.~and A.R.H.S~facilitated the collection of data from ICRAR and produced the figures.
P.J.E.~liaised with Pawsey to obtain the supercomputing data.
M.T.M.~provided data for CAS and obtained data from Keck.
All authors engaged with observatories' representatives for information. 

\section*{Competing interests}
The authors declare no competing interests.

\end{multicols}

\end{document}